\documentclass[twocolumn]{jpsj3}
\newcommand{\diff}{\mathrm{d}}
\newcommand{\imag}{\mathrm{Im}\,}

\newcommand{\trace}{\mathrm{Tr}\,}
\newcommand{\imu}{\mathrm{i}}

\title{ 
{
Electronic Order with Staggered Kondo and Crystalline Electric Field Singlets}
}
\author{Shintaro \textsc{Hoshino}\thanks{E-mail: hoshino@cmpt.phys.tohoku.ac.jp}, Junya \textsc{Otsuki} and Yoshio \textsc{Kuramoto}}

\inst{Department of Physics, Tohoku University, Sendai 980-8578}

\recdate{\today}

\abst{
{
Novel electronic order is found theoretically for a system where even number of localized electrons per site are coupled with conduction electrons.  
For precise quantitative study, a variant of the Kondo lattice model is taken with crystalline electric field (CEF) singlet and triplet states for each site.
Using the dynamical mean-field theory combined with the continuous-time quantum Monte Carlo method, 
a staggered order with alternating Kondo and CEF singlets is identified for a case 
with one conduction electron per site being distributed in 
two conduction bands each of which is quarter-filled. 
This electronic order accompanies a charge density wave (CDW) of conduction electrons that accumulate more on Kondo-singlet sites than on CEF-singlet sites.
Possible relevance of the present 
order to the scalar order in PrFe$_4$P$_{12}$ is discussed.
}
}

\kword{
Kondo singlet,
crystalline electric field singlet,
charge density wave,
dynamical mean-field theory,
continuous-time quantum Monte Carlo method,
PrFe$_4$P$_{12}$
}

\begin{document}
\maketitle

\section{Introduction}

Dichotomy between itinerant and localized characters of $f$-electrons leads to intriguing phenomena in heavy fermion systems.
With $f^1$ configuration as in Ce compounds, the behavior of the system is determined by the competition between the Kondo effect and the RKKY interaction.
The former gives itinerancy to $f$-electrons.  
If $f$ electrons are localized,
the RKKY interaction gives rise to a magnetic order.  Depending on the relative strength of the Kondo effect and the RKKY interaction,  the ground state can either be magnetic or nonmagnetic.   This picture was first spelled out by Doniach\cite{doniach77}, and is now recognized well.
On the other hand, 
in systems with two $f$-electrons per site as in some Pr and U compounds, 
the distinction between 
itinerant and localized characters is much more subtle. 
It remains 
to be
clarified how the dichotomy plays a role
in rich behaviors found in 
real systems such as PrFe$_4$P$_{12}$ and URu$_2$Si$_2$\cite{kuramoto09, hassinger08}.

A typical situation in $f^2$ systems 
is the case where the ground state is the crystalline electric field (CEF) singlet.
If $f$-electrons interact strongly with conduction electrons, the ground state can be a collective Kondo singlet 
involving
higher CEF levels.
Also in this case, $f$-electrons acquire itinerancy because of the Kondo effect.
Competition between these two singlets, namely dichotomy between itinerant and localized characters of $f$-electrons, may give rise to rich physics with exotic ordered phases.
Theoretically, there are many discussions in impurity systems with the CEF singlet\cite{shimizu95, koga96, yotsuhashi02, otsuki05, hattori05, hoshino09}.
In order to discuss electronic orders, however, investigation of periodic systems is necessary.

Let us consider the case where CEF states form
 a quasi-quartet composed of a singlet ground state and a first-excited triplet.
We represent the singlet-triplet levels 
at each site $i$
in terms of two pseudo spins
$\mib{S}_{\gamma i}$ with $\gamma=1,2$.
Correspondingly, we introduce two conduction bands $\gamma=1,2$
where electrons
interact with pseudo spins with the same orbital index $\gamma$
by the exchange interaction $J_\gamma > 0$.
Then the model reads
\begin{align}
{\cal H} & = 
   \sum _{\mib{k}\gamma \sigma} ( \varepsilon _{\mib{k}\gamma} - \mu ) c_{\mib{k}\gamma\sigma} ^\dagger c_{\mib{k}\gamma\sigma}
 + \sum _{i \gamma} J_\gamma \mib{S}_{\gamma i} \cdot \mib{s}_{{\rm c} \gamma i}    \nonumber \\
 & + 
\Delta  \sum_{i}\mib{S}_{1i} \cdot \mib{S}_{2i},
\label{eq_hamilt}
\end{align}
where 
$\mib{s}_{{\rm c}\gamma i}$ denotes the spin of conduction electrons at site $i$.
The CEF splitting $\Delta$ is simulated by
the coupling between 
$\mib{S}_{1i}$ and $\mib{S}_{2i}$
 as shown in the third term
.
We call this model the ``two-band singlet-triplet Kondo lattice model" (2BSTKLM)
in 
this paper.
This model at half filling has been investigated in one dimension by the density-matrix renormalization group\cite{watanabe99}.

In the case of $\Delta=0$, the 2BSTKLM  corresponds to the two independent Kondo lattice models\cite{lacroix79, fazekas91}.
Recently, 
a
charge density wave (CDW) 
transition 
has been
found
in the Kondo lattice model at quarter filling\cite{otsuki09}.
However, this ordered state cannot be the ground state since localized spins keep finite entropy even below the transition temperature.
In this paper, we investigate an effect of the CEF splitting on the CDW state. 
We shall demonstrate that the CEF splitting stabilizes 
a non-magnetic ordering,
and leads to a novel ground state,
which we call the ``staggered Kondo-CEF singlet order" in the following.

In order to investigate the ordering, we employ the dynamical mean-field theory (DMFT)\cite{georges96}, 
 which can be extended for the ordered state in the Kondo lattice systems.\cite{peters07, hoshino10}
The DMFT takes full account of on-site correlations, and becomes exact in infinite dimensions.
Since both Kondo and CEF effects are dominantly local,  the DMFT is suitable for our purpose.
As the impurity solver associated with the DMFT, 
we use the continuous-time quantum Monte Carlo method (CT-QMC)\cite{rubtsov05, werner06, otsuki07}.

In \S 2 we describe the formalism how to derive susceptibilities in the ordered phase with A and B sublattices
in the framework of the DMFT
.
\S 3 is devoted to technical aspects about the CT-QMC.
The main results of this paper is given in \S 4 and \S 5, where we derive
the phase diagram, and some physical quantities such as susceptibilities, order parameters, local correlation functions, and renormalized density of states.
We discuss in \S 6 how the characteristics of the system is compared with the ordinary Kondo lattice, and how the present results are relevant to understanding the scalar order in PrFe$_4$P$_{12}$.
The summary of the paper is given in \S 7, and some technical details related to \S2 are given in Appendix.

\section{Susceptibilities with Two-Sublattices}

An instability toward ordered phase is signalled by divergence of the 
corresponding susceptibility.
In this section, we generalized the DMFT formalism
\cite{georges96} so that we can
investigate the instability including two-sublattice systems.
We first consider the simplest case of non-interacting conduction electrons with  nearest-neighbor hopping in a bipartite lattice, where
the condition $\varepsilon _{\mib{k}+\mib{Q}} = - \varepsilon _{\mib{k}}$ with $\mib{Q}=(\pi, \pi, \cdots)$ is satisfied.
We use the suffix $\alpha$ for spin and orbital indices.
In particular, $\alpha = (\gamma, \sigma)$ in the 2BSTKLM given by eq.~(\ref{eq_hamilt}).
We introduce sublattice annihilation operators $c_{\mib{k}{\rm A}} (c_{\mib{k}{\rm B}})$ for A(B)-sublattice with wave vector $\mib{k}$.
The first term in eq.~(\ref{eq_hamilt})
is rewritten in terms of the sublattice operators as 
\begin{align}
{\cal H}_0 =&
{\sum_{\mib{k} \alpha}}' \left[
 \varepsilon _{\mib{k}\alpha} \left(  
 c_{\mib{k}{\rm A} \alpha}^\dagger c_{\mib{k}{\rm B}\alpha} + 
 c_{\mib{k}{\rm B} \alpha}^\dagger c_{\mib{k}{\rm A}\alpha} \right)
\right. \nonumber \\
&\left.
 - \mu \left( 
 c_{\mib{k}{\rm A}\alpha}^\dagger c_{\mib{k}{\rm A}\alpha} + c_{\mib{k}{\rm B}\alpha}^\dagger c_{\mib{k}{\rm B}\alpha} \right)
\right]  ,
\end{align}
where the summation $\sum '$ is taken over
wave vectors $\mib{k}$ belonging to the reduced Brillouin zone with sublattices, {\it i.e.},
half of the original Brillouin zone.

The effect of terms other than ${\cal H}_0$ in eq.(\ref{eq_hamilt}) appears as the self-energy 
$\Sigma _{\lambda \alpha}(z)$ 
of conduction electrons. 
where $\lambda = {\rm A, B}$ indicates the label for the sublattice.
Note that the self energy is spatially local in the DMFT.
In terms of the quantity
$\zeta_{\lambda \alpha} (z) = z + \mu - \Sigma _{\lambda \alpha}(z)$,
the full Green function is given by
\begin{align}
& G^{\lambda \lambda'}_{ \mib{k} \alpha} (z) = \nonumber \\
& \frac{1}{\zeta _{\rm A\alpha}(z) \zeta _{\rm B \alpha}(z) - \varepsilon_{\mib{k}\alpha}^2}\  [\  \zeta _{\bar \lambda \alpha}(z) \delta_{\lambda \lambda'} +  \varepsilon_{\mib{k}\alpha} (1-\delta_{\lambda \lambda'})\ ]
. \label{eq_g_lattice}
\end{align}
where ${\bar \lambda}$ indicates the complementary component such as ${\bar {\rm A}} = {\rm B}$.

Let us assume a symmetric density of states: $\rho (\varepsilon ) = (N/2)^{-1}\sum_{\mib{k}}' \delta (\varepsilon - \varepsilon _{\mib{k}}) = \rho (- \varepsilon )$.
Note that $\rho (\varepsilon)$ is the same as the density of states in the original Brillouin zone.
In the symmetric case, the local Green function becomes diagonal with respect to the sublattice label.
This is easily seen from eq.(\ref{eq_g_lattice}) where the off-diagonal part is an odd function of $\varepsilon_{\mib{k}\alpha}$, and vanishes by 
summation over $\mib{k}$.
Then the local Green function is given by 
\begin{align}
G^{\lambda }_{ {\rm loc}, \alpha} (z) = \frac{1}{N/2}\  {\sum_{\mib{k}}}' G^{\lambda \lambda}_{ \mib{k} \alpha} (z).
\label{eq_g_loc}
\end{align}

Magnetic and charge susceptibilities can be evaluated from the two-particle Green function, which is given in the imaginary time domain by
\begin{align}
&\chi^{\lambda \lambda'}_{\alpha \alpha'} (\tau_1, \tau_2, \tau_3, \tau_4) = \nonumber \\ 
&\frac{1}{N/2}\  {\sum _{\mib{k},\mib{k}'} }' \left[ 
\langle T_\tau c^\dagger _{\mib{k}{\lambda} \alpha}(\tau_1)c _{\mib{k}\lambda \alpha}(\tau_2) c^\dagger _{\mib{k}' \lambda'\alpha'}(\tau_3)c _{\mib{k}' \lambda'\alpha'}(\tau_4) \rangle \right. \nonumber\\ 
&\left. -  \langle T_\tau c^\dagger _{\mib{k} \lambda\alpha}(\tau_1)c _{\mib{k}\lambda\alpha}(\tau_2) \rangle 
          \langle T_\tau c^\dagger _{\mib{k}'\lambda' \alpha'}(\tau_3)c _{\mib{k}' \lambda' \alpha'}(\tau_4) \rangle \right]  
 .  \label{eq_two_part}
\end{align}
Here we only consider the uniform component in the reduced Brillouin zone.
The uniform and staggered components in the original Brillouin zone can be calculated by the linear combinations as
\begin{align}
\chi^{\rm unif, stag} = \frac{1}{2} (\chi^{\rm AA} + \chi^{\rm BB} \pm 2\chi^{\rm AB})
. \label{eq_sublattice_sum}
\end{align}
The Fourier transform for the two-particle Green function (\ref{eq_two_part}) is defined by
\begin{align}
\chi ({\rm i}\varepsilon _n , {\rm i}\varepsilon _{n'}; {\rm i}\nu _m)  =  
\frac{1}{\beta ^2} \int _0 ^\beta {\rm d} \tau _1 {\rm d} \tau _2 {\rm d} \tau _3 {\rm d} \tau _4 \nonumber \\
\times
 e^{{\rm i}\varepsilon _n (\tau _2 -\tau _1)} e^{{\rm i}\varepsilon _{n'} (\tau _4 -\tau _3)} e^{{\rm i}\nu _m (\tau _2 -\tau _3)}  
\chi (\tau_1, \tau_2, \tau_3, \tau_4)  
, \label{eq_fourier}
\end{align}
where $\varepsilon _n$ and $\nu _m$ are the fermionic and bosonic Matsubara frequencies, respectively.
The susceptibility which represents a response to external fields is derived from $\chi ({\rm i}\varepsilon _n , {\rm i}\varepsilon _{n'}; {\rm i}\nu _m)$ as
\begin{align}
\chi_{\alpha \alpha'}^{\lambda \lambda'} (\imu \nu _m) = 
\frac{1}{\beta} \sum _{nn'} \chi_{\alpha \alpha'}^{\lambda \lambda'} (\imu \varepsilon_n, \imu \varepsilon_{n'} ; \imu \nu _m).
\end{align}

\begin{figure}
\begin{center}
\includegraphics[width=85mm]{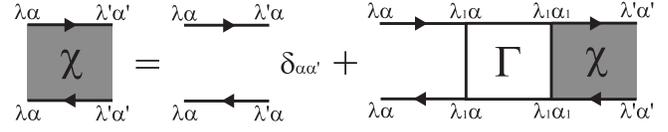}
\caption{
Bethe-Salpeter equation for the two-particle Green function in the two-sublattice formulation. 
Summation should be taken
over the sublattice index $\lambda_1$ and the spin-orbital
index $\alpha _1$.
}
\label{fig_bs_eq}
\end{center}
\end{figure}

We use the Bethe-Salpeter equation to relate $\chi_{\alpha \alpha'}^{\lambda \lambda'}$ to the effective impurity model.
Noting that the vertex part is local and now depends on the sublattice, we obtain the following equation:
\begin{align}
&\chi^{\lambda \lambda'}_{\alpha \alpha'} ({\rm i}\varepsilon _n , {\rm i}\varepsilon _{n'}, {\rm i}\nu _m) 
= 
\chi^{0, \lambda \lambda'}_{ \alpha} ({\rm i}\varepsilon _n ; {\rm i}\nu _m) \delta_{\alpha \alpha'} \delta_{nn'} \nonumber \\
+ &\sum_{\lambda_1 \alpha_1 n_1} \chi^{0, \lambda \lambda_1}_{\alpha} ({\rm i}\varepsilon _{n} ; {\rm i}\nu _m)  \nonumber \\
\times &\Gamma^{\lambda _1}_{\alpha \alpha_1}({\rm i}\varepsilon _{n} , {\rm i}\varepsilon _{n_1} ; {\rm i}\nu _m) \chi^{\lambda_1 \lambda'}_{\alpha_1 \alpha'} ({\rm i}\varepsilon _{n_1} , {\rm i}\varepsilon _{n'} ; {\rm i}\nu _m) 
 . \label{eq_bs_lattice}
\end{align}
The function $\chi^0$ is the two-particle Green function without the vertex part, and defined by
\begin{align}
\chi^{0,\lambda \lambda'}_{\alpha} ({\rm i}\varepsilon _n ; {\rm i}\nu _m) 
= - \frac{1}{N/2} {\sum _{\mib{k}}}' 
G_{\mib{k} \alpha} ^{\lambda \lambda'} (\imu \varepsilon _n)
G_{\mib{k} \alpha} ^{\lambda' \lambda} (\imu \varepsilon _n + {\rm i}\nu _m)
. \label{eq_free_chi_lattice}
\end{align}
Equation~(\ref{eq_bs_lattice}) is graphically shown in Fig. \ref{fig_bs_eq}.

We evaluate the local vertex $\Gamma^{\lambda}$ using the effective impurity model.
We define the local two-particle Green function as
\begin{align}
&\chi^{\lambda }_{{\rm loc}, \alpha \alpha'} (\tau_1, \tau_2, \tau_3, \tau_4) = \nonumber \\ 
&\langle T_\tau c^\dagger _{{\lambda} \alpha}(\tau_1)c _{\lambda \alpha}(\tau_2) c^\dagger _{ \lambda\alpha'}(\tau_3) c _{ \lambda\alpha'}(\tau_4) \rangle  \nonumber\\ 
 &-  \langle T_\tau c^\dagger _{ \lambda\alpha}(\tau_1)c _{    \lambda\alpha}(\tau_2) \rangle 
          \langle T_\tau c^\dagger _{ \lambda \alpha'}(\tau_3)c _{ \lambda \alpha'}(\tau_4) \rangle   
, \label{eq_two_part_loc}
\end{align}
where $c_{\lambda \alpha} = (N/2)^{-1/2}\sum _{\mib{k}} '  c_{\mib{k} \lambda \alpha}$ is the annihilation operator at the origin site for each sublattice.
In the effective impurity model, the Bethe-Salpeter equation for $\chi^{\lambda }_{{\rm loc}, \alpha \alpha'}$ is expressed by local quantities as 
\begin{align}
&\chi^{\lambda }_{{\rm loc}, \alpha \alpha'} ({\rm i}\varepsilon _n , {\rm i}\varepsilon _{n'}, {\rm i}\nu _m) 
= 
\chi^{0, \lambda }_{{\rm loc},  \alpha} ({\rm i}\varepsilon _n ; {\rm i}\nu _m) \delta_{\alpha \alpha'} \delta_{nn'} \nonumber \\
&+ \sum_{ \alpha_1 n_1} \chi^{0, \lambda }_{{\rm loc}, \alpha} ({\rm i}\varepsilon _{n} ; {\rm i}\nu _m)  \nonumber \\
&\times \Gamma^{\lambda}_{\alpha \alpha_1}({\rm i}\varepsilon _{n} , {\rm i}\varepsilon _{n_1} ; {\rm i}\nu _m) 
\chi^{\lambda }_{{\rm loc}, \alpha_1 \alpha'} ({\rm i}\varepsilon _{n_1} , {\rm i}\varepsilon _{n'} ; {\rm i}\nu _m) 
 , \label{eq_bs_loc}
\end{align}
where 
\begin{align}
\chi^{0,\lambda }_{{\rm loc}, \alpha} ({\rm i}\varepsilon _n ; {\rm i}\nu _m) 
= - 
G_{{\rm loc}, \alpha} ^{\lambda} (\imu \varepsilon _n)
G_{{\rm loc}, \alpha} ^{\lambda} (\imu \varepsilon _n + {\rm i}\nu _m)
.
\end{align}
We note that, in general, the local two-particle Green function in eq.~(\ref{eq_two_part_loc}) has the off-diagonal element such as $\chi_{\rm loc}^{\rm AB}$.
{
However with the symmetric property $\rho (\varepsilon ) = \rho (- \varepsilon )$, the off-diagonal elements vanish because
$G_{\rm loc}^{\lambda \lambda'}$ is diagonal.
}

In the DMFT, the vertices in eqs.~(\ref{eq_bs_lattice}) and (\ref{eq_bs_loc}) are the same.
Hence, we can obtain $\chi^{\lambda \lambda'}$ by solving these equations simultaneously.
To make the notation simple, we use a matrix form with respect to $(n, \alpha)$.
Then eq. (\ref{eq_bs_loc}) is rewritten
as
\begin{align}
[\chi_{\rm loc}^{\lambda} ]
^{-1}
=
[\chi_{\rm loc}^{0,\lambda} ]
^{-1}
- 
\Gamma ^{\lambda}.
\end{align}
Similarly, eq.~(\ref{eq_bs_lattice}) is rewritten as
\begin{align}
\begin{pmatrix}
\chi^{\rm AA} & \chi^{\rm AB} \\
\chi^{\rm BA} & \chi^{\rm BB} 
\end{pmatrix}
^{-1}
= 
\begin{pmatrix}
\chi^{0,\rm AA} & \chi^{0,\rm AB} \\
\chi^{0,\rm BA} & \chi^{0,\rm BB} 
\end{pmatrix}
^{-1}
- 
\begin{pmatrix}
\Gamma ^{\rm A} &  \\
 & \Gamma ^{\rm B} 
\end{pmatrix}
.
\end{align}
Consequently,
the problem is reduced to the calculation of the local two-particle Green function $\chi_{\rm loc}^{\rm A}$ and $\chi_{\rm loc}^{\rm B}$
in the effective impurity model.
These can be evaluated by the two-particle $t$-matrix as discussed in ref. \citen{otsuki09-2}.
The details of the calculations of
$\chi^{0}$ and $\chi^{0}_{\rm loc}$ are given in Appendix.

\section{CT-QMC Algorithm}
In the DMFT, a periodic model is mapped to an effective impurity model.\cite{georges96}
Here we explain how to apply the CT-QMC to the present impurity model.
An impurity version of
the 2BSTKLM (\ref{eq_hamilt}) is written as follows:
\begin{align}
{\cal H} &= {\cal H}_{\rm c} + {\cal H}_f + {\cal H}_{\rm int}
, \label{eq_imp_ham}  \\
{\cal H}_{\rm c} &= \sum _{\mib{k}\gamma \sigma} \varepsilon _{\mib{k}\gamma}
 c_{\mib{k}\gamma\sigma} ^\dagger c_{\mib{k}\gamma\sigma} 
 + \sum_{\gamma\sigma} v_\gamma c_{\gamma \sigma}^\dagger c_{\gamma \sigma}
, \\
{\cal H}_f &= - \Delta P_{\rm s} + \frac{\Delta}{4}
\label{eq_imp_ham_f}
, \\
{\cal H}_{\rm int} & = \sum _{\gamma \sigma \sigma'} \frac{J_\gamma}{2} X^\gamma_{\sigma \sigma'} (c_{\gamma \sigma'}^\dagger c_{\gamma \sigma} - \alpha_\gamma \delta_{\sigma \sigma'}) + \sum_{\gamma}\alpha_\gamma \frac{J_\gamma}{2} ,
\label{eq_imp_ham_int}
\end{align}
where $v_\gamma = - J_\gamma /4$, and $X^\gamma_{\sigma \sigma'} = | \gamma \sigma \rangle  \langle \gamma \sigma' |$ is the $X$-operator for the localized states, 
and $P_{\rm s} = - \mib{S}_1 \cdot \mib{S}_2 + 1/4$ is the projection operator onto the CEF singlet state.
The parameter $\alpha _\gamma$ is chosen as 0 for ferromagnetic interaction ($J_\gamma <0$) and 1 for antiferromagnetic coupling ($J_\gamma >0$) in order to avoid 
the minus sign problem as noted in ref. \citen{hoshino09}.
The constant term may be neglected 
in the simulation.

The partition function 
$Z$
is expanded 
with respect to ${\cal H}_{\rm int}$
using the formula
\begin{align}
Z = \trace \left\{ T_{\tau} e^{-\beta {\cal H}_0} \exp \left[ - \int_{0}^{\beta} \diff \tau {\cal H}_{\rm int}^{\rm I} (\tau) \right] \right\} , \label{partition}
\end{align}
where the non-interacting Hamiltonian ${\cal H}_0$ is defined by ${\cal H}_0 = {\cal H}_{\rm c}+{\cal H}_f$.
The suffix `${\rm I}$' denotes the interaction picture: ${\cal H}_{\rm int}^{\rm I} (\tau) = e^{\tau {\cal H}_0} {\cal H}_{\rm int} e^{ - \tau {\cal H}_0}$.
The CT-QMC evaluates the perturbation expansion by Monte Carlo method\cite{rubtsov05, werner06}.
The trace over conduction electrons is computed with the aid of Wick's theorem. 
For the localized 
spins, 
on the other hand, we evaluate the following expression:
\begin{eqnarray}
W_f = \langle X^{\gamma _k {\rm I}}_{\sigma _{k} \sigma _{k} '}(\tau _{k}) \cdots X^{\gamma _1 {\rm I}}_{\sigma _1 \sigma _1 '}(\tau _1) \rangle _f 
\label{eq_loc_conf}
\end{eqnarray}
Here 
$k$ is the perturbation order, and
we consider the imaginary-time set with $\beta > \tau _k  > \cdots > \tau _1 \geq 0$.
The average is defined by $\langle \cdots \rangle _f = {\rm Tr}_f [e^{-\beta {\cal H}_f} \cdots] / Z_f$ with $Z_f = 3 + e^{\beta \Delta}$.
The efficient algorithm for the Kondo model~\cite{otsuki07,hoshino09} using the ``segment"
cannot be applied to the present model, 
because the localized state $| \gamma \sigma \rangle$ is not an eigenstate of ${\cal H}_f$.
Therefore, we must multiply the $k$ matrices to evaluate eq.~(\ref{eq_loc_conf}).

We rewrite eq.~(\ref{eq_loc_conf}) as
\begin{align}
W_f &= \frac{1}{Z_f} {\rm Tr}_f \left[ { F}_k { F}_{k-1} \cdots { F_1} \right] 
, \label{eq_mat_mul} \\
{ F}_i &= \rho(\tau_{i+1} - \tau_i)  X^{\gamma _i}_{\sigma _{i} \sigma _{i} '}
 ,
\end{align}
where we have introduced $\rho (\tau) = e^{-\tau {\cal H}_f}$, and
$\tau _{k+1}$ is defined by $\tau _{k+1} = \beta + \tau _1$.
This multiplication process takes the main part of the computational time, and the calculation becomes heavy compared to the method using segments.
This difficulty can be somewhat reduced by using the so-called tree algorithm or binning algorithm\cite{computation}.
Although the direct multiplication takes the computing time of $O(k)$, the tree algorithm and binning algorithm give $O(\log k)$ and $O(\sqrt{k})$, respectively.
We have implemented the binning algorithm, since it is much simpler than the other.

For evaluating eq. (\ref{eq_mat_mul}), it is convenient to choose the basis that diagonalizes $\rho (\tau)$.
We denote this eigenstates
by $(|{\rm s} \rangle, |{\rm t}+ \rangle, |{\rm t}0 \rangle, |{\rm t}- \rangle)$, and $\rho (\tau)$ becomes
\begin{eqnarray}
\rho (\tau)
=
\begin{pmatrix}
e^{\tau \Delta} & & & \\
 & 1& & \\
 & & 1 & \\
 & & & 1 
\end{pmatrix}
.
\end{eqnarray}
Correspondingly, the $X$-operator $X_{\sigma \sigma'}^\gamma$ expressed in the basis
($|\uparrow \uparrow \rangle$, $|\uparrow \downarrow \rangle$, $| \downarrow \uparrow \rangle$, $| \downarrow \downarrow \rangle$)
is replaced by
$U X_{\sigma \sigma'}^\gamma U^{-1}$,
where
\begin{eqnarray}
U = 
\begin{pmatrix}
0&1/\sqrt{2}&-1/\sqrt{2}&0 \\
1&0&0&0 \\
0&1/\sqrt{2}&1/\sqrt{2}&0 \\
0&0&0&1 \\
\end{pmatrix}
 = {^{\rm t}} (U ^{-1}) .
\end{eqnarray}
Using this representation, 
it can be shown that negative weight does not appear
up to the second order of $J_\gamma$.
The absence in fact persists to higher orders empirically. 
In the presence of magnetic fields, however, the negative sign may arise.

Next we discuss how to calculate physical quantities.
The algorithm for conduction electrons is the same as the previous CT-QMC methods.
For localized part, the quantity $\langle T_\tau A(\tau) B \rangle$ with $A$ and $B$ being operators for localized states is given by the following expression:
\begin{align}
&\langle T_\tau A(\tau) B \rangle = \nonumber \\
\frac{1}{Z_f} &\left< 
{\rm Tr}_f \left[ { F}_k \cdots F_{i+1} A^{\rm I}(\tau - \tau_{i+1}) F_i \cdots { F_1} B \right]  \right> _{\rm MC}
,
\end{align}
where $\langle \cdots \rangle _{\rm MC}$ means the Monte Carlo average, and we have assumed $\tau \in [\tau_i , \tau_{i+1}]$.
With use of this formula, we can evaluate quantities such as $\left< \mib{S}_1 \cdot \mib{S}_2 \right>$ and time-dependent correlation functions.
We note that there is another method to calculate the correlation functions using the inverse matrix of the Green function for conduction electrons\cite{otsuki07}.

The correlation between localized and conduction spins can be evaluated by using another method.
We write the Hamiltonian as ${\cal H} = {\cal H}_0 + \lambda {\cal H}_{\rm int}$, and the partition function as $Z_\lambda$.
With the expansion
$Z_\lambda = \sum _{k=0} ^{\infty} \lambda ^k Z_k$, we obtain the following equation:
\begin{align}
 \left.  \frac{1}{Z_\lambda} \frac{\partial Z_\lambda}{\partial \lambda} \right| _{\lambda = 1} 
= \frac{1}{Z} \sum _{k=0} ^{\infty} k Z_k \equiv \left< k \right> _{\rm MC} 
\label{eq_deriv}
\end{align}
From eq. (\ref{partition}), on the other hand, the left-hand side of (\ref{eq_deriv}) is given by $- \beta\left<  {\cal H}_{\rm int} \right> $.
Thus we derive the formula
\begin{align}
\langle {\cal H}_{\rm int} \rangle = - \left<  k \right> _{\rm MC} /\beta
. \label{eq_form_corr}
\end{align}
Since the Hamiltonian ${\cal H}_{\rm int}$ includes the interaction between conduction and localized spins, we can evaluate $\left< \mib{S}_\gamma \cdot \mib{s}_{\rm c\gamma} \right>$ from this expression.
Note that the expression (\ref{eq_form_corr}) cannot be applied to the time-dependent correlation functions.

\begin{figure}
\begin{center}
\includegraphics[width=85mm]{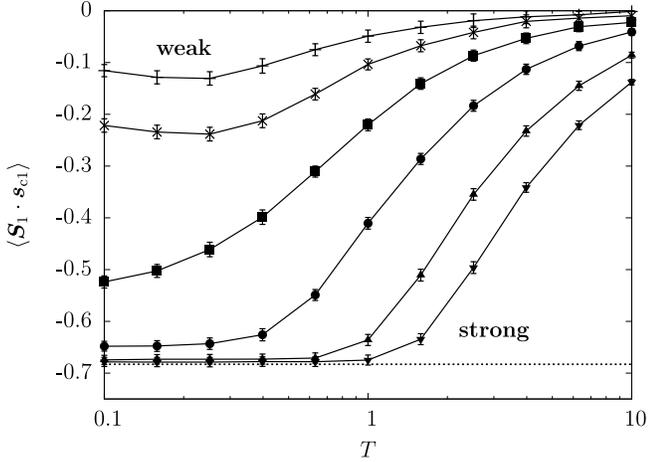}
\caption{
Equal-time correlation $\left< \mib{S}_1 \cdot \mib{s}_{\rm c1} \right>$ as a function of temperature.
The parameters are $J_1 = J_2 = \Delta = 0.5,\  1,\  2,\  4,\  8,\  12$ from top to bottom.
The dotted line shows the value $\left< \mib{S}_1 \cdot \mib{s}_{\rm c1} \right> \simeq-0.683$ for the ground state in the strong coupling limit .
}
\label{fig_scl}
\end{center}
\end{figure}

At the end of this
section, we show an exemplary result of our simulation to check the accuracy.
Figure \ref{fig_scl} shows the equal-time correlation $\left< \mib{S}_1 \cdot \mib{s}_{\rm c1} \right>$ as a function of temperature
 with $J_1 = J_2 = \Delta$.
Here we have used the rectangular density of states $\rho_0 (\varepsilon) = \theta (D - |\varepsilon| ) / 2D$, and put $D=1$.
In the strong-coupling limit $J_{\gamma}, \Delta \gg D$,
the impurity model given by eq. (\ref{eq_imp_ham})
is reduced to four-spin system since the kinetic energy term can be neglected.
Diagonalizing this Hamiltonian, we obtain 
$\left< \mib{S}_1 \cdot \mib{s}_{\rm c1} \right> = -(1+\sqrt{3})/4 
\simeq
-0.683$ at $T=0$.
The result in Fig.~\ref{fig_scl} tends to this value, and 
$\left< \mib{S}_1 \cdot \mib{S}_2 \right>$ tends to $-1/4$.
Even though the CT-QMC is 
based on 
the expansion from the weak coupling, we have confirmed that it can reproduce the strong coupling limit.

\section{Phase Diagram}
In the rest of this paper, 
we put $\varepsilon_{\mib{k}1} = \varepsilon_{\mib{k}2} = \varepsilon_{\mib{k}}$,
and use a tight-binding band on a hypercubic lattice. 
The density of states is given by
\begin{eqnarray}
\rho_0 (\varepsilon) = \frac{1}{D} \sqrt{\frac{2}{\pi}} \exp \left[ -2 \left( \frac{\varepsilon}{D} \right) ^2 \right]
,
\end{eqnarray}
with $D=1$ as a unit of energy.
This band  has a perfect nesting property with the wave vector $\mib{Q}$ at half filling\cite{georges96}.
In order to discuss the effect of the CEF splitting, we fix the strength of the interaction as $J_1 = J_2 = J = 0.8$ and vary the CEF splitting $\Delta$.
Note that the definition of $J$ is different from that in ref. \citen{otsuki09} by factor 2.
We also fix the number of conduction electrons per site as $n_{\rm c} = 1$, which corresponds to the quarter filling of both conduction bands.

\subsection{Divergence of  Susceptibilities}

\begin{figure}
\begin{center}
\includegraphics[width=85mm]{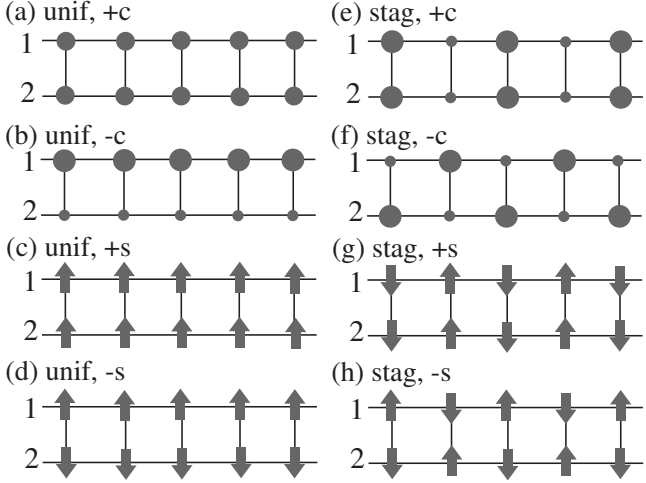}
\caption{
Illustration of electronic orders probed by divergence of susceptibilities.
The labels 1 and 2 are the orbital indices of conduction electrons.
}
\label{fig_phys_img}
\end{center}
\end{figure}

To determine the phase diagram, we search for instability of normal states in terms of 
divergent response of
conduction electrons.
The magnetic and charge response functions are defined by
\begin{align}
\chi_{\rm c, s} = \frac{1}{2} ( \chi_{\uparrow\uparrow} + \chi_{\downarrow\downarrow} \pm 2 \chi_{\uparrow\downarrow} ), 
\end{align}
where the indices `c' and `s' indicate charge and spin channel, respectively.
In a similar manner, we define `$+$' and `$-$' channels from the orbital index
\begin{align}
\chi_{\pm} = \frac{1}{2} ( \chi_{11} + \chi_{22} \pm 2 \chi_{12} ) .
\end{align} 
Combining with the spatial dependence defined in eq.~(\ref{eq_sublattice_sum}), we consider the following eight susceptibilities:
$\chi _{\pm {\rm c}}^{\rm unif}, \chi _{\pm{\rm s}}^{ \rm unif}, \chi _{\pm{\rm c}}^{\rm stag}, \chi _{\pm {\rm s}}^{\rm stag}$.
{
Figure \ref{fig_phys_img} illustrates the electronic orders in the strong coupling limit, each of which is probed by the corresponding susceptibility.
}

\begin{figure}
\begin{center}
\includegraphics[width=85mm]{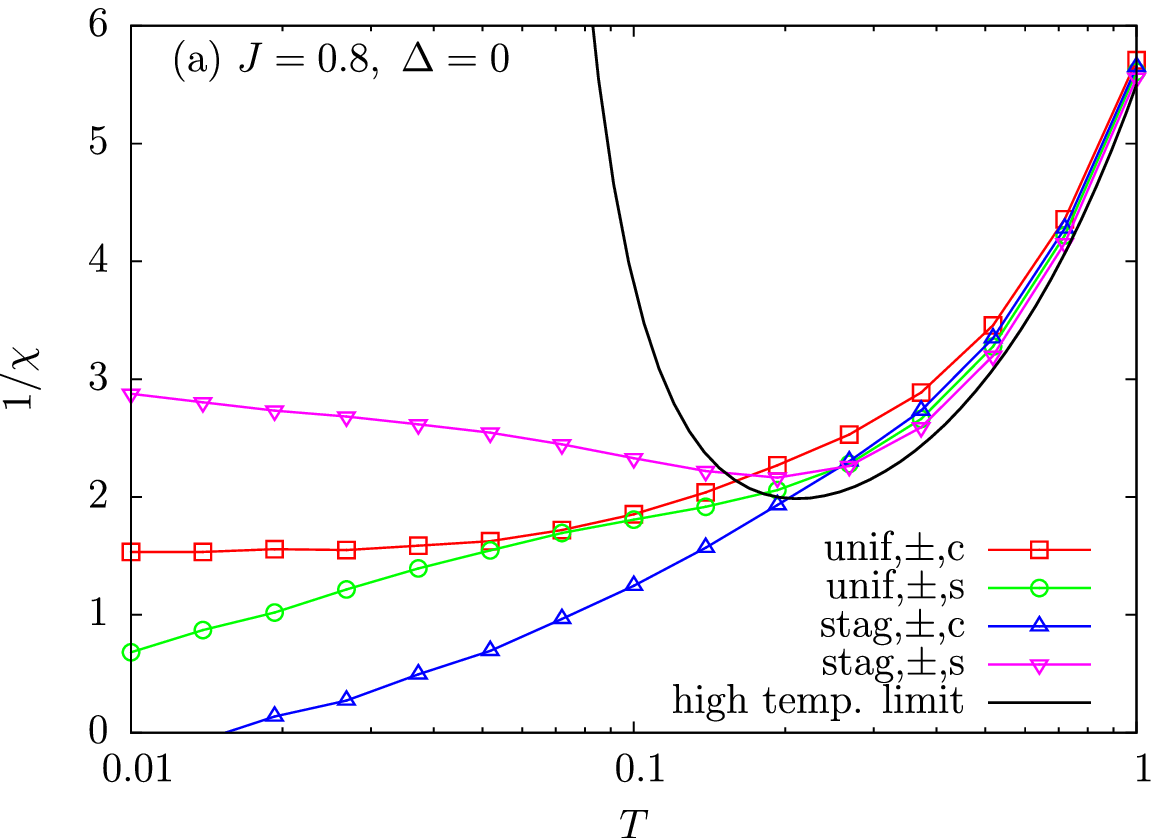}
\includegraphics[width=85mm]{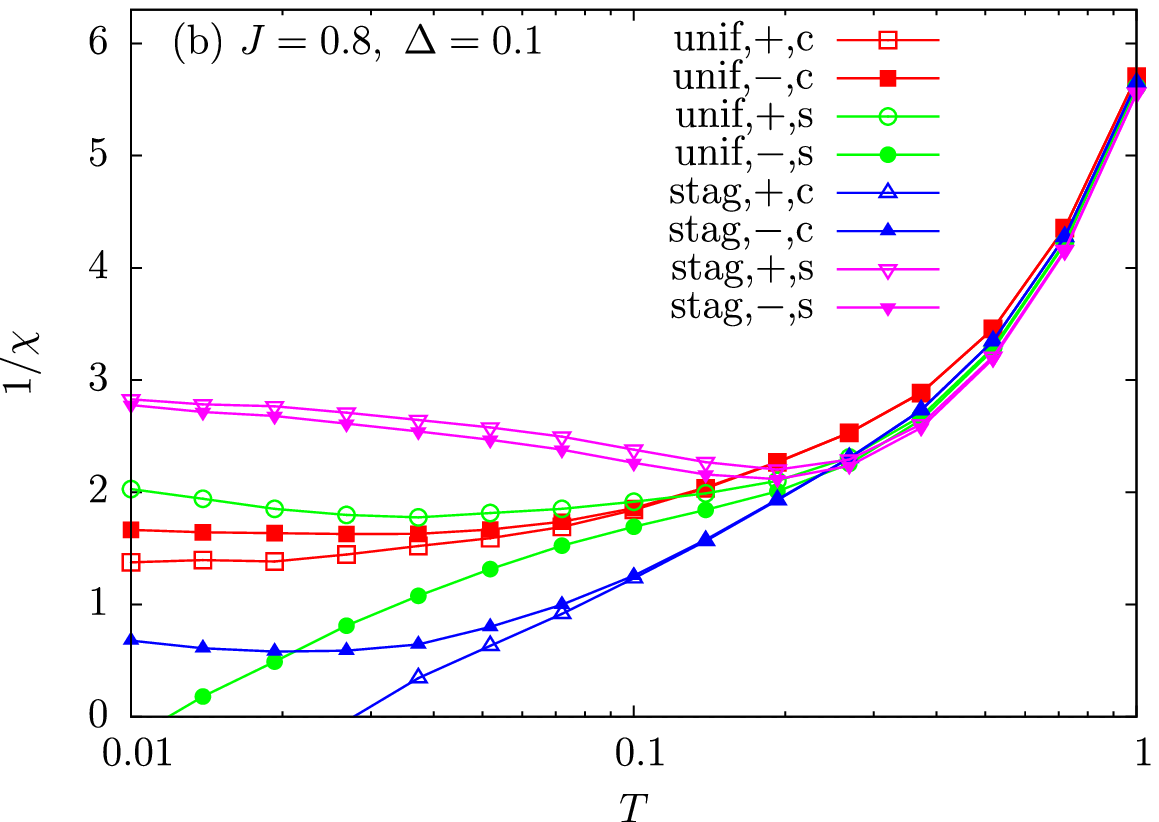}
\caption{
(color online)
Inverse susceptibilities for (a) $\Delta = 0$ and (b) $\Delta = 0.1$.
In (a), $+$ and $-$ channels are equivalent since there is no off-diagonal component
$\chi_{12}$.
The high-temperature behavior is given in eq.~(\ref{eq_high_temp}).
}
\label{fig_suscep_all}
\end{center}
\end{figure}

Figure \ref{fig_suscep_all} shows the temperature dependence of 
inverse susceptibilities for
$\Delta = 0$ and $\Delta = 0.1$
.
Note that the case with $\Delta = 0$
shown in Fig.~\ref{fig_suscep_all}(a)
corresponds to the pair of ordinary
Kondo lattice models at quarter filling.
In this case, $\chi _+$ and $\chi _-$ are equivalent because of $\chi_{12}=0$.
For comparison, we include in 
Fig.~\ref{fig_suscep_all}(a) 
the high temperature form of the susceptibility 
given by
\begin{align}
\chi_{\rm high} = \frac{1}{\beta} \sum _{n} \left( \frac{1}{\imu \varepsilon _n + \mu _0} \right) ^2 = \beta f(\mu_0) f(-\mu_0 )
, 
\label{eq_high_temp}
\end{align}
where $f(x) = 1 / (e^{\beta x} + 1)$ is the Fermi distribution function and $\mu _0$ is the chemical potential without interaction.  
The susceptibilities of the 2BSTKLM 
show good agreement with $\chi _{\rm high}$ in the high temperature region.

In Fig. \ref{fig_suscep_all}(a), 
$\chi _{\pm \rm c}^{\rm stag}$ tends to diverge at $T = T_{\rm CDW} \sim 0.015$, which indicates the CDW instability.
In this ordered phase, two states illustrated in Figs.~\ref{fig_phys_img}(e) and (f) are degenerate.
This degeneracy is resolved by the singlet-triplet splitting.
The CEF splitting suppresses $\chi_{-{\rm c}}^{\rm stag}$ and enhances 
$\chi_{+{\rm c}}^{\rm stag}$ leading to a higher transition temperature.
Figure~\ref{fig_suscep_all}(b) demonstrates this situation, giving 
$T_{\rm CDW}\sim 0.028$.
As will be shown in \S 5.1, this CDW state is identified as the staggered Kondo-CEF singlet order.
Although $\chi_{-{\rm s}}^{\rm unif}$ also diverges at $T\sim0.012$, this is not meaningful because
the ordered state probed by
$\chi_{-{\rm s}}^{\rm unif}$ 
should be the reference state below $T_{\rm CDW}$.

\begin{figure}
\begin{center}
\includegraphics[width=85mm]{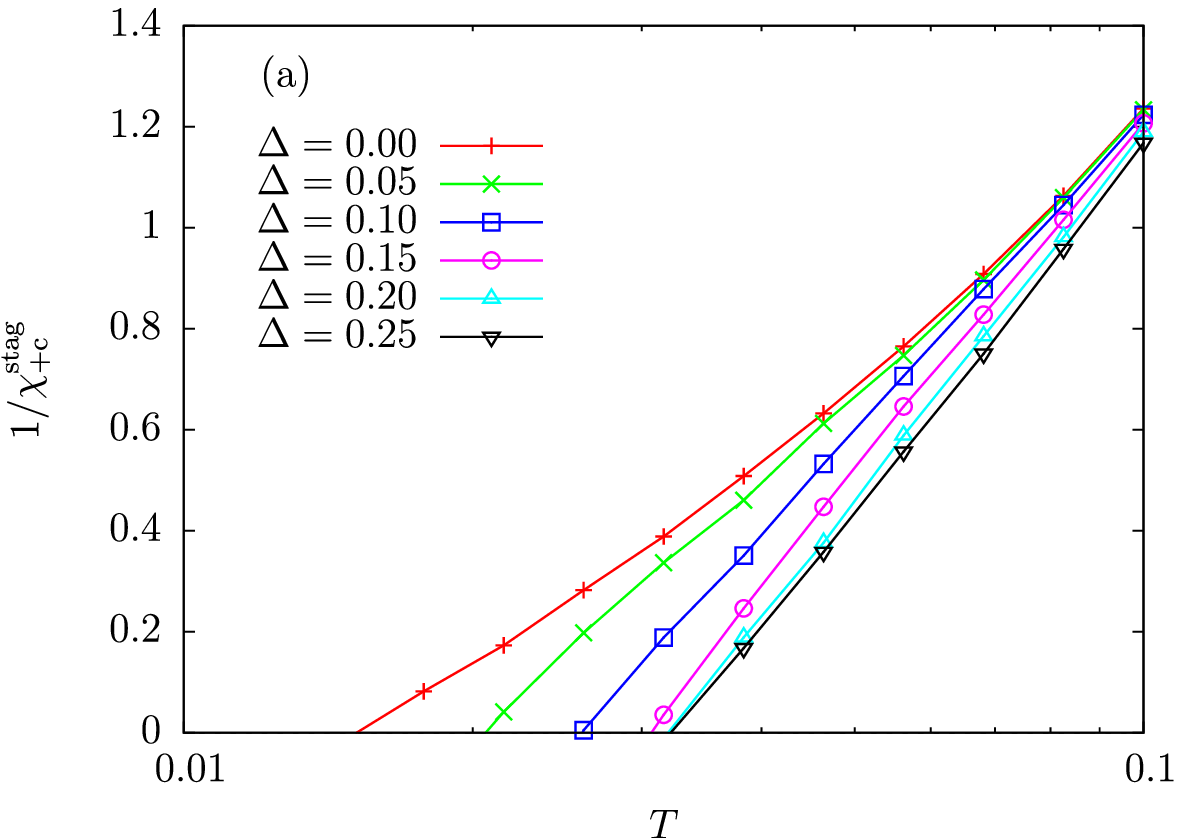}
\includegraphics[width=85mm]{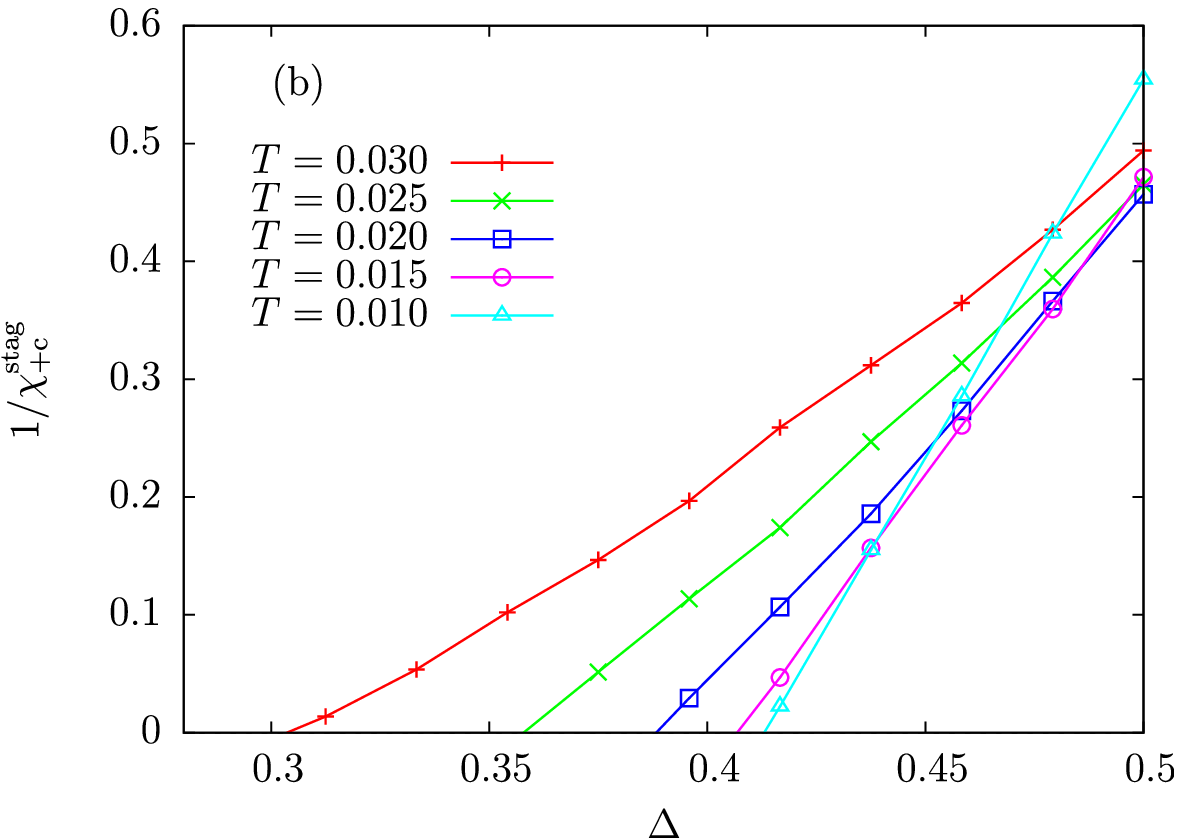}
\caption{
(color online)
Inverse susceptibility $1/\chi_{+{\rm c}}^{\rm unif}$ as functions of (a) temperature $T$ and (b) CEF splitting $\Delta$.
}
\label{fig_suscep_d_depend}
\end{center}
\end{figure}

\begin{figure}
\begin{center}
\includegraphics[width=85mm]{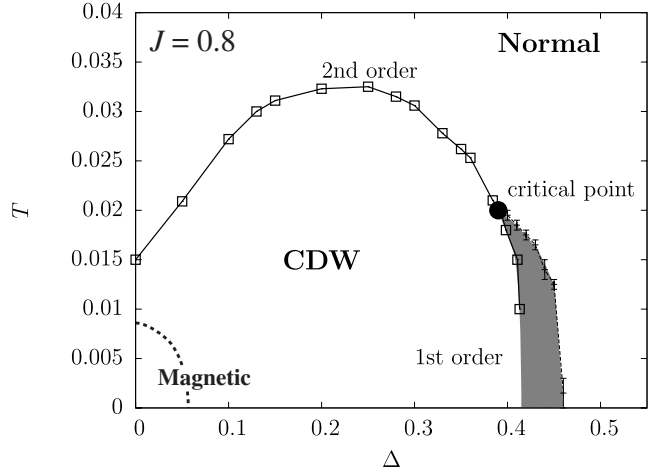}
\caption{
Phase diagram of the 2BSTKLM with $J=0.8$ and one conduction electron per site.
The phase labelled ``CDW'' is identified as the staggered Kondo-CEF singlet order.
The shaded area shows the region of first-order transition, which ends at
the critical point.  See text for details of derivation.
Magnetic ordering is expected with small $\Delta$ and $T$ as indicated by ``Magnetic''.
}
\label{fig_phase_d}
\end{center}
\end{figure}

Let us now focus on $\chi_{+{\rm c}}^{\rm stag}$ that has been found to diverge.
Figure \ref{fig_suscep_d_depend}(a) shows the temperature dependence of 
$\chi_{+{\rm c}}^{\rm stag}$
for
various values of $\Delta$.
The transition temperature $T_{\rm CDW}$ increases with increasing 
$\Delta$.
For $\Delta > 0.3$, on the other hand, it is 
{ more convenient}
 to 
consider the susceptibility as a function of 
$\Delta$
with fixed $T$ 
because of the sharp decrease of $T_{\rm CDW}$.
Fig. \ref{fig_suscep_d_depend}(b) shows that the critical value of $\Delta$ increase with decreasing temperature.
As clearly seen from this figure, the transition points have almost no difference between $T = 0.015$ and $T=0.01$.
Then we conclude that there is no ordering for $\Delta \gtrsim 0.41$.

In this way,
we obtain the $T$-$\Delta$ phase diagram as shown in Fig. \ref{fig_phase_d} for $J=0.8$ with one conduction electron per site.
The increase of $T_{\rm CDW}$ in the small $\Delta$ region indicates 
a 
stabilization of the
staggered Kondo-CEF singlet order
by the CEF splitting.
Sufficiently large $\Delta$ destroys the order, and the system becomes normal state with the CEF singlet.
The transition changes from the second order to the first order around $\Delta = 0.4$, which will be 
discussed in the next subsection.
For sufficiently small $\Delta$,
we expect a magnetically ordered ground state, as in the ordinary Kondo lattice,
because of the residual entropy of the localized spins.
This region is indicated qualitatively in Fig. \ref{fig_phase_d} as ``Magnetic''.
More details about this aspect will be discussed in \S 6.1.
We have also taken other values of coupling constant such as $J=0.6$ or $1.0$, and found no qualitative change in the overall behavior.

\subsection{Growth of Order Parameter and Hysteresis}

\begin{figure}
\begin{center}
\includegraphics[width=85mm]{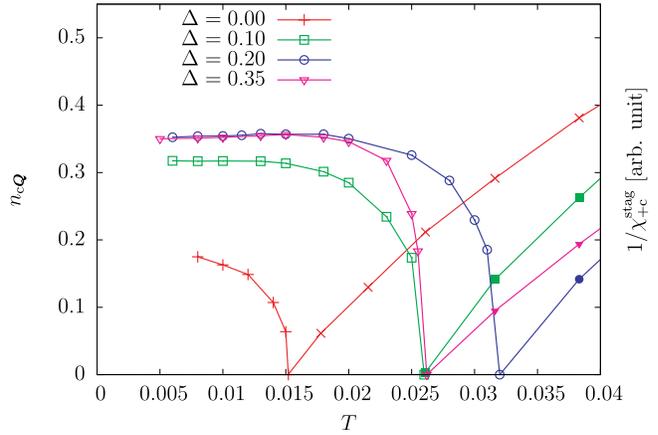}
\caption{
(color online)
Temperature dependence of the order parameter $n_{\rm c\mib{Q}}$ (left scale) as a function of temperature.
Also shown is the inverse susceptibility $[\chi_{+\rm c}^{\rm stag}]^{-1}$ (right scale).
}
\label{fig_moment}
\end{center}
\end{figure}

We discuss the order parameter defined by $n_{{\rm c}\mib{Q}} = (n_{\rm cA}- n_{\rm cB})/2$.
Figure \ref{fig_moment} shows the temperature dependence of 
$n_{{\rm c}\mib{Q}}$.
{
The transition temperature $T_{\rm CDW}$, which
is determined by divergence of the susceptibility, is also shown in Fig. \ref{fig_moment}.
The order parameter grows as $(T_{\rm CDW}-T)^{1/2}$ 
as in the usual mean-field theory.
We note that the initial rise around the transition temperature becomes sharper with increasing $\Delta$.
}
\begin{figure}
\begin{center}
\includegraphics[width=85mm]{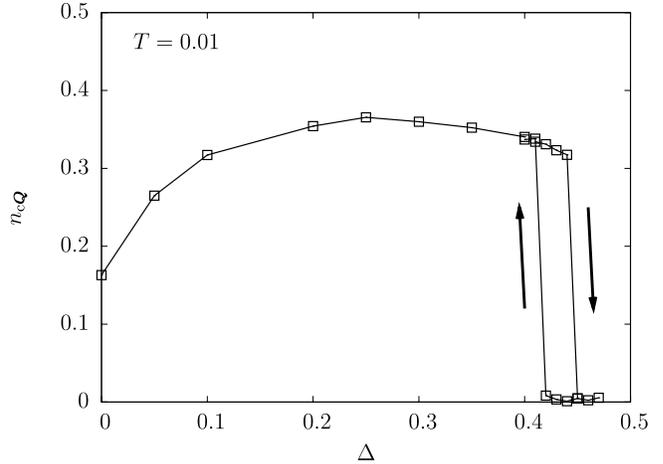}
\caption{
Order parameter vs $\Delta$ with fixed temperature at $T=0.01$.
The arrows show the hysteresis that occurs
depending on whether we approach from smaller or larger $\Delta$.
}
\label{fig_hysteresis}
\end{center}
\end{figure}

Next we show the presence of first-order transition
by deriving
the order parameter as a function of $\Delta$ at fixed temperature.
Figure \ref{fig_hysteresis} shows the result at $T=0.01$.
In the region with small $\Delta$, 
$n_{{\rm c}\mib{Q}}$ 
becomes larger for larger CEF splitting, which indicates stabilization of the 
staggered Kondo-CEF singlet order.
This behavior corresponds to the increase of $T_{\rm CDW}$ shown in Fig. \ref{fig_phase_d}.
The magnitude of $n_{\rm c\mib{Q}}$ becomes the maximum at $\Delta \sim 0.25$, and slowly decreases with larger splitting.

As shown in Fig. \ref{fig_hysteresis},  we have observed the hysteresis 
around $\Delta \sim 0.42$.
This is a characteristic for the first-order transition.  
When approaching from smaller $\Delta$, 
we have used the initial condition with a staggered chemical potential
for the effective medium of the DMFT.
On the other hand, when approaching from larger CEF splitting,
common chemical potential is used for both sublattices.
The shaded area in Fig. \ref{fig_phase_d} shows the region 
of hysteresis.
Note that the hysteresis is observed only in the region with $T \lesssim 0.022$.
This means the existence of the critical point 
from second- to first-order transitions.
More details are discussed in terms of the susceptibility in the next subsection.

\subsection{Critical Point from First- to Second-Order Transitions}
\begin{figure}
\begin{center}
\includegraphics[width=85mm]{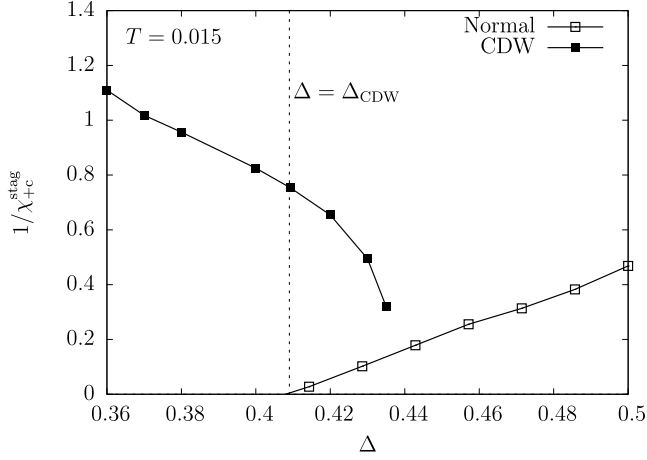}
\caption{
Inverse susceptibility with temperature fixed at $T=0.015$ both inside and outside the ordered phase.
With decreasing $\Delta$ in the disordered phase,
$1/\chi_{+{\rm c}}^{\rm stag}$ tends to zero at $\Delta = \Delta_{\rm CDW}$, which is obtained by extrapolation.
While $1/\chi_{+{\rm c}}^{\rm stag}$ 
derived in the ordered phase
is finite at $\Delta = \Delta_{\rm CDW}$
.
}
\label{fig_suscep_first}
\end{center}
\end{figure}

We study the hysteresis in more detail near $\Delta \sim 0.4$. 
Figure \ref{fig_suscep_first} shows
the inverse susceptibility at $T=0.015$  in both disordered and ordered
phases.
For calculation of the susceptibility in the ordered phase,  the formulation given in \S 2 is used. 
If the transition is of second order, both susceptibilities must diverge at the same transition 
point
.
The result in Fig.~\ref{fig_suscep_first} shows divergent
susceptibility in the normal state 
 at $\Delta = \Delta_{\rm CDW}$, while
the 
staggered Kondo-CEF singlet order
phase has the finite susceptibility at this point.
Hence this difference clearly shows the first-order nature of the transition.
It is difficult to reach the divergence of the susceptibility from the 
staggered Kondo-CEF singlet order phase
since tiny statistical errors 
destroy the meta-stable ordered states.

\begin{figure}
\begin{center}
\includegraphics[width=85mm]{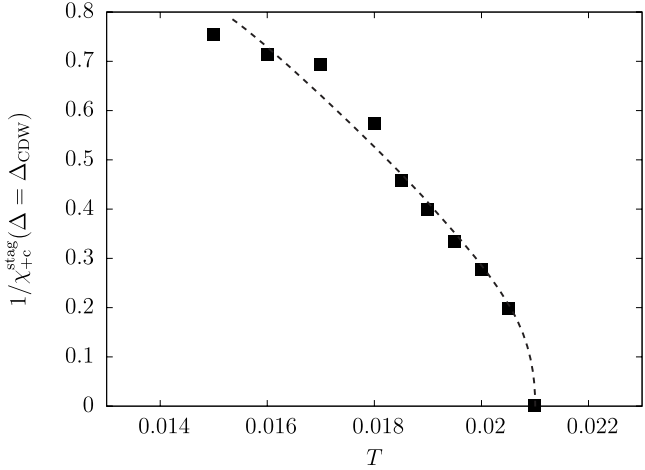}
\caption{
Inverse susceptibility in the ordered phase as a function of temperature.
The value $\Delta$ at each temperature is taken to be $\Delta_{\rm CDW}$ as explained in the text.
The dotted line is guided to the eye.
}
\label{fig_suscep_critical}
\end{center}
\end{figure}

Now let us derive the critical point where the transition changes from second to first order. 
Following the procedure similar to that shown in Fig.~\ref{fig_suscep_first}, 
we derive $\Delta_{\rm CDW}$ 
for a given temperature 
by extrapolating 
$1/\chi_{+{\rm c}}^{\rm stag}$ to zero from large-$\Delta$ side.
Then we calculate $\chi_{+{\rm c}}^{\rm stag}$ at $\Delta=\Delta_{\rm CDW}$
in the ordered phase taking the same temperature. 
Figure \ref{fig_suscep_critical} shows the result of 
$\chi_{+{\rm c}}^{\rm stag}$ calculated in this way
for different values of temperature.
The apparent scattering of the data for $T\lesssim 0.018$
comes mainly from ambiguity of extrapolation of the inverse susceptibility.
With increasing temperature, the susceptibility becomes larger, and tends to diverge 
at $T = T_{\rm cr} \simeq 0.021$ with $\Delta = \Delta_{\rm cr} \simeq 0.39$.
Since the susceptibility in the ordered phase must diverge also at $\Delta_{\rm CDW}$ if the transition is of second-order, we conclude that $T_{\rm cr}$
is the critical point where the character of the order changes from first to second order.
The critical point is shown by the black circle in Fig. \ref{fig_phase_d}.

The change from second- to first-order transitions can be qualitatively explained by the Landau theory\cite{chaikin}.  The free energy is then given by
\begin{align}
f = \frac{r}{2} \phi ^2 + b \phi ^4 + c \phi ^6
,  \label{eq_landau6}
\end{align}
where $\phi$ is an order parameter, and the coefficient $c$ must be positive.
The coefficient $r$ is given by
$r= a(T-T_{\rm c})$ where $T_{\rm c}$ is a second-order transition temperature.
The character of the transition is of second order for $b > 0$, but becomes of first order for $b < 0$.
The case with $b = 0$ corresponds to the critical point.
The Landau theory 
also
explains the temperature dependence of order parameters shown in Fig. \ref{fig_moment}.
At the critical point with $b=0$, $\phi$ near the transition point behaves as $\phi \sim r^{1/4}$, where the critical exponent is $1/4$ instead of $1/2$.
Therefore if $b$ approaches as $b \rightarrow +0$, the initial rise becomes sharper reflecting the change of the exponent as in Fig. \ref{fig_moment}.

\section{Correlation Functions and Density of States}

\subsection{Local Correlation Functions}

\begin{figure}
\begin{center}
\includegraphics[width=85mm]{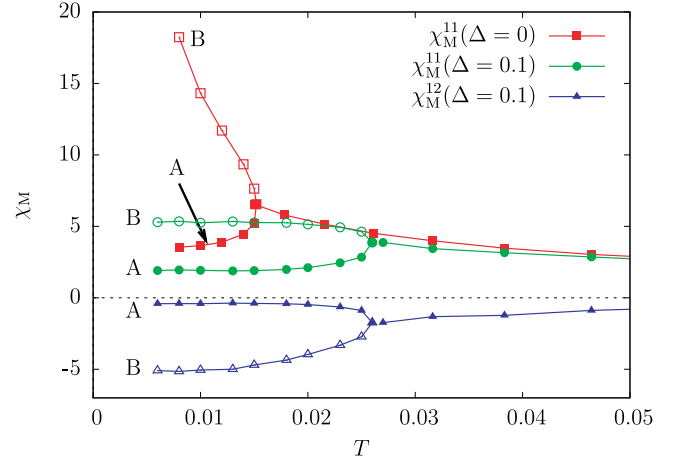}
\caption{
(color online)
{
Local magnetic susceptibilities $\chi^{\gamma \delta}_{\rm M}$
vs temperature.
With $\Delta=0$, the system decomposes into two independent Kondo lattices,
and the off-diagonal component $\chi^{12}_{\rm M}$ vanishes.
The labels A and B show the electron-rich and electron-poor sites, respectively.
}}
\label{fig_local_suscep}
\end{center}
\end{figure}

In this section, we elucidate the nature of the ordered state 
from correlation functions.
Using pseudo spins, the local magnetic susceptibilities are defined by
\begin{align}
\chi_{\rm M}^{\gamma \delta} = \int _{0} ^{\beta}  \left[ \langle T_\tau S_\gamma^z (\tau) S_\delta^z \rangle 
 -  \langle S_\gamma^z \rangle  \langle S_\delta^z \rangle \right]  \diff \tau
.
\end{align}
Since the magnetic dipole is given by $J_z = \sum_{\gamma} a_\gamma S_\gamma ^z$
with appropriate $a_\gamma$, the magnetic susceptibility is represented as $\chi_J = \sum_{\gamma \delta} a_\gamma a_\delta \chi_{\rm M}^{\gamma \delta}$.
The parameter $a_\gamma$ depends on the wave functions 
of the singlet-triplet
states\cite{shiina04}.
Let us discuss the local magnetic susceptibility with $\Delta = 0$ shown in Fig. \ref{fig_local_suscep}, which corresponds to 
a pair of
the Kondo lattice models.
In this case, the relations $\chi_{\rm M}^{11} = \chi_{\rm M}^{22}$ and $\chi_{\rm M}^{12} = 0$ are satisfied.
Below
the transition temperature $T\sim 0.015$, $\chi_{\rm M}^{11}$ splits into two values because of the emergence of the staggered CDW order.
{ 
The B-sublattice has conduction electrons fewer than the A-sublattice.
Then 
the B-sublattice
shows a Curie-like behavior reflecting the 
localized spins with $\Delta = 0$.
The remaining entropy of the localized spins may lead to the magnetic instability at low temperatures.
On the other hand, the susceptibility for A-sublattice is strongly suppressed.
This clearly indicates the formation of the Kondo singlet 
at
A-sublattice.
}

\begin{figure}
\begin{center}
\includegraphics[width=85mm]{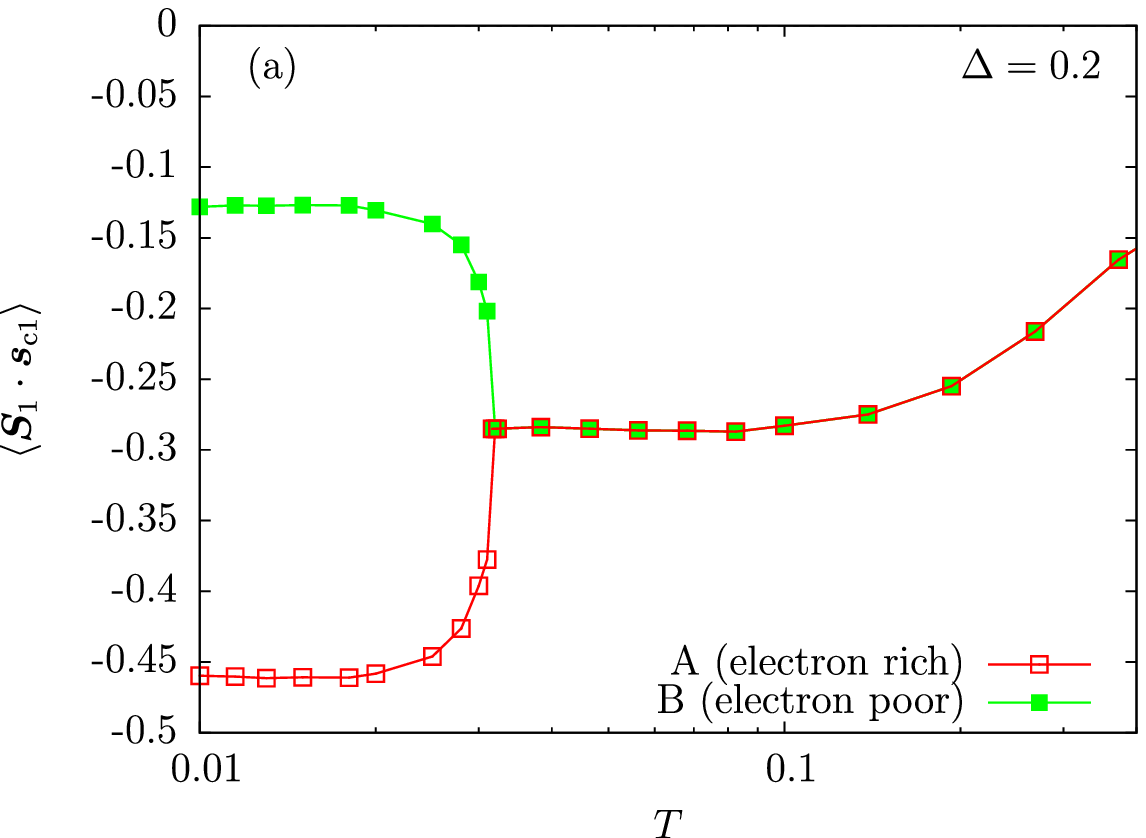}
\includegraphics[width=85mm]{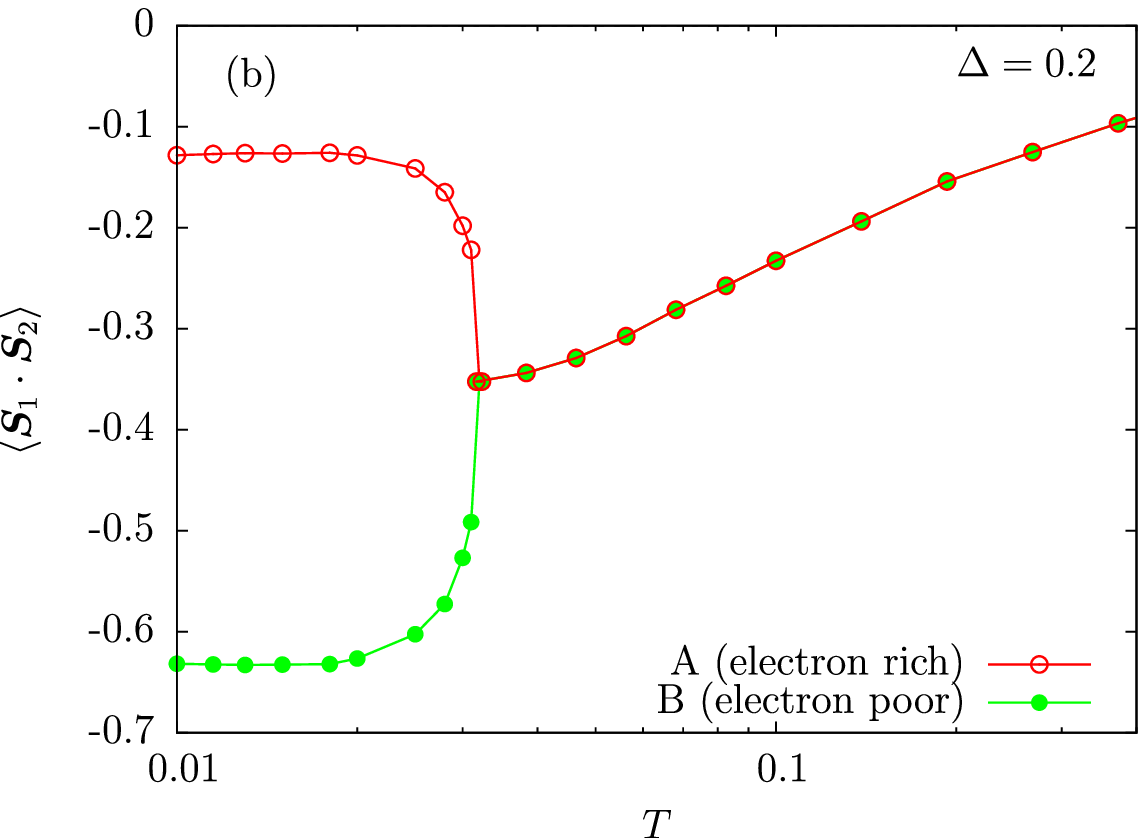}
\caption{
(color online)
Local equal-time correlations (a) $\langle \mib{S}_1 \cdot \mib{s}_{\rm c1} \rangle$ and (b) $\langle \mib{S}_1 \cdot \mib{S}_2 \rangle$ with 
$\Delta = 0.2$.
}
\label{fig_local_corr}
\end{center}
\end{figure}

\begin{figure}
\begin{center}
\includegraphics[width=85mm]{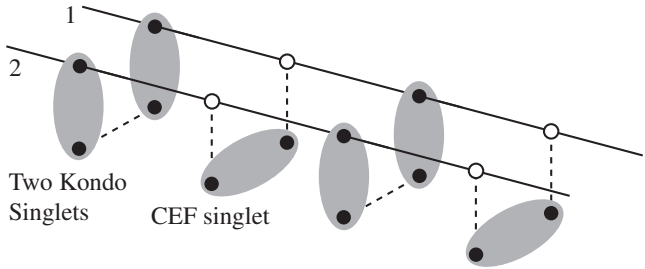}
\caption{
{
Schematic picture for the staggered order with Kondo and CEF singlets.
The lower part represents localized electrons, while the upper part on the lines represents conduction electrons.
}
The black and white circles show the presence and absence of electrons, respectively.
}
\label{fig_sing_sing}
\end{center}
\end{figure}

Next we consider the situation with the finite CEF splitting
also shown in Fig. \ref{fig_local_suscep}.
In the case 
of
$\Delta = 0.1$, 
$\chi_{\rm M}^{12}$ is finite owing to the correlation between pseudo spins as shown in Fig. \ref{fig_local_suscep}.
The spin fluctuation at B-sublattice is suppressed, and all susceptibilities
 show paramagnetic behavior.

Let us examine equal-time spin correlations that
clarify properties of each sublattice
with the finite CEF splitting.
We first consider 
$\langle \mib{S}_1 \cdot \mib{s}_{\rm c1} \rangle$, which is equal to $\langle \mib{S}_2 \cdot \mib{s}_{\rm c2} \rangle$ under the present condition.
Figure \ref{fig_local_corr}(a) shows the temperature dependence of
$\langle \mib{S}_1 \cdot \mib{s}_{\rm c1} \rangle$, which
is enhanced 
on
A-sublattice and suppressed 
on
B-sublattice.
Bearing $n_{\rm cA} > n_{\rm cB}$ in mind, we conclude that the localized spin on A-sublattice forms the Kondo singlet.
On the other hand, the correlation $\langle \mib{S}_1 \cdot \mib{S}_2 \rangle$ between localized spins is enhanced at B site as shown in Fig. \ref{fig_local_corr}(b).
Hence, the B-sublattice corresponds to the CEF singlet.

Let us estimate
the magnitude of the effective CEF splitting for the CEF-singlet site.
Taking the effective Hamiltonian for the localized states as ${\cal H}_f^{\rm eff} = {\tilde \Delta} \mib{S}_1 \cdot \mib{S}_2$,
we obtain 
the susceptibility as $\chi_{\rm M}^{12} = - 1 / (2{\tilde \Delta})$ for the ground state.
Here ${\tilde \Delta}$ is the effective CEF splitting, 
which can be estimated in B-sublattice 
using the value in Fig. \ref{fig_local_suscep}.
The result is ${\tilde \Delta} = 0.098$, which is very close to the original CEF splitting $\Delta = 0.1$.
Hence,  spatially extended Kondo singlets do not significantly affect the magnitude of the CEF splitting.
Besides, in Figure \ref{fig_local_corr}(b), $\langle \mib{S}_1 \cdot \mib{S}_2 \rangle$ in the low-temperature limit is not far from $-0.75$ expected for the isolated singlet.
Hence the CEF singlet
is almost decoupled from conduction electrons.
We note that the correspondence between $\Delta$ and $\tilde \Delta$ does not hold for $\Delta \lesssim 0.05$.
This corresponds to the fact that the CEF-singlet site tends to be magnetically polarized
near $\Delta = 0$.

Thus, the present order with one conduction electron per site turns out to be 
a staggered order with the Kondo and CEF singlets.
Figure \ref{fig_sing_sing} schematically shows this staggered order.
Except for the strong coupling limit, the number of conduction electrons at the CEF singlet site is not zero because the Kondo singlets are spatially extended.

\subsection{Density of States}

The single-particle dynamics can be derived from the Green function (\ref{eq_g_lattice}) and (\ref{eq_g_loc}).
The density of states of conduction electrons is given by
\begin{align}
\rho (\omega) = \frac{1}{2} \sum_{\lambda={\rm A,B}} \left[ - \frac{1}{\pi} \imag G^{\lambda}_{\rm loc} (\omega + \imu \delta) \right]
.
\end{align}
We note that the Green functions 
does not depend
on the labels $\gamma$ and $\sigma$ in the present condition.
The Pad\'{e} approximation is used 
{
for analytic continuation from imaginary Matsubara frequencies.
}

\begin{figure}
\begin{center}
\includegraphics[width=85mm]{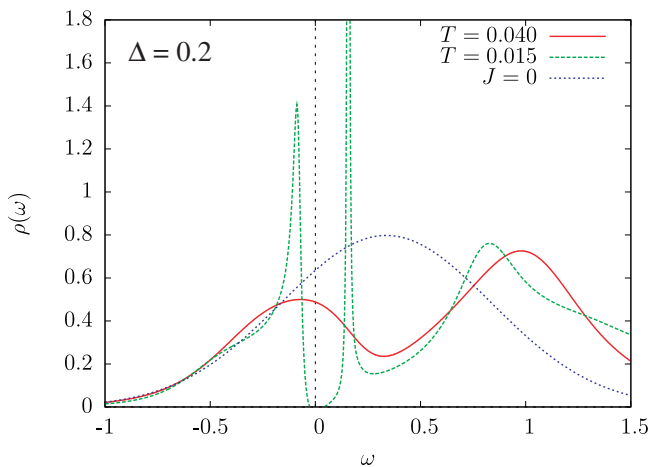}
\caption{
(color online)
{
Density of states of conduction electrons
in normal ($T=0.015$) and ordered ($T=0.010$) phases
with $\Delta = 0.2$. 
The bare density of states ($J=0$) at quarter filling is also shown for comparison.
}
}
\label{fig_dos}
\end{center}
\end{figure}

Figure \ref{fig_dos} shows the density of states 
for $\Delta = 0.2$.
In the disordered state ($T=0.04$),
the density of states shows the metallic behavior.
{
The pseudo-gap structure around $\omega \sim 0.3$ is interpreted as a kind of hybridization
 between conduction electrons and the localized states by the Kondo effect.
Although there is no real hybridization because the pseudo spins do not have charge degrees of freedom, strong renormalization by the Kondo effect gives rise to an electronic state that allows this interpretation. 
}
In the ordered state ($T=0.015$), an energy gap opens at the Fermi level. 
The sharp peak below the gap comes from the sublattice for the Kondo singlet, while the peak above the gap is due to the CEF singlet site. 
Hence, this 
double-peak structure 
clearly shows the difference of 
the occupation number
between the Kondo and CEF singlet sites.

The origin of the insulating behavior is explained as follows.
Each band has one conduction electron per unit cell in the ordered state.
Provided that the localized spin at the Kondo-singlet site participates to the conduction band, each band is filled by two ``electrons''.
Then the system can be an insulator as in the Kondo lattice at half filling.
In this viewpoint, the staggered Kondo-CEF singlet order may be regarded as 
alternating itinerant and localized sites of $f$-electrons.

\section{Discussion}
\subsection{Relation to CDW in Ordinary Kondo Lattice}

We discuss how the new electronic order found in the present paper is related to known orders in the ordinary Kondo lattice.
In the limit of $\Delta =0$, the present 2BSTKLM is reduced to a pair of 
Kondo lattice models.
Let us briefly summarize the electronic order found in the ordinary Kondo lattice.
In addition to the magnetic order,
it has been found in ref. \citen{otsuki09} that the Kondo lattice has 
a CDW order at quarter filling.
In the strong coupling,
the CDW is visualized as alternating Kondo singlets and 
localized spins.  
The corresponding electron number $n_{\rm c}$ of conduction electrons
per site is 1/2.
Associated with hopping of conduction electrons, 
which is regarded as perturbation from the strong-coupling limit,
a Kondo singlet and a local spin can exchange their positions.
This process is of first-order with respect to hopping.
On the other hand, the second-order perturbation leads to inter-site attraction between a Kondo singlet and a local spin\cite{hirsch84,sigrist}.
Although smaller attraction arises
between Kondo singlets as well, 
the total second-order perturbation 
gives effective inter-site repulsion between Kondo singlets.
Hence, 
this repulsive interaction gives a chance to stabilize the CDW order by partially sacrificing the hopping energy with intermediate coupling.
In infinite dimensions, the CDW order is indeed stabilized at quarter filling with $n_{\rm c}= 0.5$ according to ref. \citen{otsuki09}.

Since the uncompensated spin sites still have substantial entropy, the CDW alone cannot be the ground state in the Kondo lattice.
It is likely that remaining spins form a magnetic order on top of the CDW background \cite{otsuki09}.
The magnetic fluctuation is clearly seen in the local magnetic susceptibility as shown in Fig. \ref{fig_local_suscep} with $\Delta = 0$.
Unless the CEF splitting is large enough, this situation may 
remain
in the 2BSTKLM with finite $\Delta$.
The region where we expect the magnetic ground state is roughly drawn in Fig. \ref{fig_phase_d}.
However, we have not been able to demonstrate its existence, since the solution with small $\Delta$ and $T$ in the two-sublattice DMFT does not converge.
This indicates that the magnetic order has a longer periodicity than described by the two-sublattice system.

Let us now consider the ordered phase in the 2BSTKLM with 
larger
$\Delta$.
We assume that the strong coupling limit in the 2BSTKLM is described by Kondo-singlet site and CEF-singlet site.
In this case, the entropy vanishes even without magnetic order.
One may regard the Kondo-singlet site as occupied by a fictitious spinless fermion, and the CEF singlet site as vacant site of the fermion, {\it i.e.}, a hole.
In a similar manner to the ordinary Kondo lattice in the strong coupling limit, the spinless fermions have an effective hopping and inter-site repulsion.
This inter-site interaction tends to form a non-magnetic order, namely the staggered Kondo-CEF singlet order.
It is clear that this non-magnetic order 
is a characteristic of non-Kramers systems.
As shown in Figs. \ref{fig_phase_d} and \ref{fig_hysteresis},
moderate values of $\Delta$ stabilize the staggered Kondo-CEF singlet order.

\subsection{Itinerant and Localized Characters}

Next we discuss the staggered Kondo-CEF singlet order from the aspect of itinerant and localized characters of $f$-electrons.
For $f^1$ system, the itinerant character is realized by the Kondo effect
 as heavy fermion state.   
 If $f$ electrons are localized, on the contrary, a magnetic order appears 
by the RKKY interaction.
The competition between the Kondo effect and the RKKY interaction
leads to the quantum phase transition between the ordered and disordered phases.
For $f^2$ system with CEF singlet, 
both the itinerant and localized limits are disordered phases where the Kondo effect is dominant in the itinerant regime.
The staggered Kondo-CEF singlet order 
is realized in the competing region, and
interpreted as alternating sites of  itinerant and localized states of $f$-electrons.
In the weak-coupling, the itinerant character is responsible for the insulating ground state at quarter filling.

The inter-site interaction leading to 
the staggered Kondo-CEF singlet order is different from the
RKKY interaction.
The effective repulsion between the Kondo singlets is the dominant mechanism for 
the present order.
It is notable that the RKKY interaction is understood from the weak coupling limit, while the 
present staggered Kondo-CEF singlet order 
is understood naturally from the strong coupling limit.

\subsection{Relevance to Real Systems}

Let us finally discuss possible application of the present results to understanding 
PrFe$_4$P$_{12}$.
This material shows the Kondo-like behavior in the resistivity and undergoes a non-magnetic order at $T=6.5{\rm K}$\cite{aoki05}.
In the ordered phase, a field-induced staggered moment is observed\cite{iwasa08-2}.
From phenomenological and experimental analysis, this order is identified as a scalar order\cite{kiss06, sakai, kikuchi07}, but the corresponding microscopic state is not yet clear.
Inelastic neutron scattering experiment shows characteristic behaviors such as broad quasi-elastic peak in the disordered phase, and the inelastic peak in the ordered phase\cite{iwasa03, iwasa08, park08}.

We remark that in 
the present staggered Kondo-CEF singlet order, 
the difference of 
the local susceptibility
$\chi_{\rm M}$ between two sublattices results in an appearance of field-induced antiferromagnetic moment.  
Furthermore, our model naturally explains the appearance of CEF excitations only below the transition temperature.
However, the realistic band structure\cite{sugawara00}
is rather different from the identical two conduction bands taken in the present paper, which leads to an insulating ground state.
More refinement is necessary for serious comparison with real systems, which
will be given in separate publications.

\section{Summary and Outlook}

We have applied the DMFT combined with CT-QMC to the 2BSTKLM  where CEF singlet-triplet states interact with two-band conduction electrons.
The instability of the staggered ordered phase is derived by using the 
formulation of the susceptibility in two-sublattice systems.
In the framework of the DMFT, 
physical quantities such as susceptibility, order parameter, correlation functions and the density of states have been calculated at finite temperatures.

In the 2BSTKLM  with one conduction electron per site, we have 
found the staggered order with Kondo and CEF singlets.
The equal-time correlation shown in Fig. \ref{fig_local_corr} clearly shows this staggered ordering.
This electronic order accompanies the CDW of conduction electrons because they gather at the Kondo singlet site 
to screen the localized moments.
Below the transition temperature, the system becomes insulating as in the Kondo insulator, which is seen in the density of states.
With different character of conduction bands, however, the insulating behavior should no longer hold.

Although we have considered only magnetic and charge susceptibilities in this paper, it is also possible to calculate the pairing susceptibility for $s$-wave superconductivity in the DMFT.
In the 2BSTKLM, a pairing is possible mediated by the CEF excitation from singlet to triplet states.
Indeed, we have observed in preliminary calculations 
an instability toward superconductivity at low temperatures. 
Here the singlet pairing between electrons in different conduction bands is realized.
We shall discuss
aspects related to the superconductivity in a separate paper.

\acknowledgement
The authors are grateful to K. Iwasa for the fruitful discussions.
One of the authors (S. H.) is supported by the global COE program of MEXT Japan.
This work was partly supported by a Grand-in-Aid for Scientific Research on Innovative Areas "Heavy Electrons" (No 20102008) of The Ministry of Education, Culture, Sports, Science, and Technology, Japan.

\appendix
\section{Useful Formulae for Susceptibilities}
As we have seen in \S 2, the susceptibility in the lattice system is derived from the local susceptibility.
In addition to the calculation of the local susceptibility, we need to evaluate the susceptibilities without the vertex part.
In this Appendix, we derive relevant formulae to calculate the local susceptibility.
First of all, we define the following two complex functions:
\begin{align}
F_1(z) &= \int \diff \varepsilon \frac{\rho(\varepsilon)}{z^2 - \varepsilon^2}
= \frac{g(z)}{z}  ,\\
F_2(z) &= \int \diff \varepsilon \frac{\rho(\varepsilon)}{(z - \varepsilon)^2}
= - \frac{\diff g(z)}{\diff z} ,
\end{align}
where
\begin{align}
g(z) = \int \diff \varepsilon \frac{\rho(\varepsilon)}{z - \varepsilon}.
\end{align}
We have used the symmetric condition $\rho (\varepsilon) = \rho ( - \varepsilon)$.
As shown later, the functions $F_1$ and $F_2$ are related to the staggered and uniform components, respectively.
We can calculate the local Green function in the original Brillouin zone from $g(z)$.
In the case with the hypercubic lattice, for example, $g(z)$ is represented by an error function\cite{georges96}.

The local Green function in the two-sublattice system is given by eq. (\ref{eq_g_loc}).
Only the diagonal elements survive the summation with respect to $\mib{k}$,
and are written as
\begin{align}
G^{\lambda }_{\rm loc} (z) =  \zeta _{\bar \lambda} (z)  \  F_1 \left({\tilde z} \right) ,
\end{align}
with ${\tilde z} = \sqrt{\zeta _{\rm A} (z)\zeta _{\rm B} (z)}$.
We omit the suffix $\alpha$ throughout this Appendix.
On the other hand, the uniform susceptibility given in eq. (\ref{eq_free_chi_lattice}) is calculated from the following form:
\begin{align}
&\frac{1}{N/2} { \sum_{\mib{k}} }' G^{\rm AA}_{ \mib{k}} (z_1) G^{\rm AA}_{ \mib{k}} (z_2) \nonumber \\
 = &\frac{\zeta_{\rm B}(z_1)  \zeta_{\rm B}(z_2)}{{\tilde z}^2_1 - {\tilde z}^2_2} [ F_1({\tilde z}_2)- F_1({\tilde z}_1)]
\label{eq_formula_1} \\
&\frac{1}{N/2} { \sum_{\mib{k}} }' G^{\rm AB}_{ \mib{k}} (z_1) G^{\rm BA}_{ \mib{k}} (z_2)
 = \frac{ {\tilde z}^2_2 F_1({\tilde z}_2) - {\tilde z}^2_1 F_1({\tilde z}_1) }{{\tilde z}^2_1 - {\tilde z}^2_2}
\label{eq_formula_2}
\end{align}
In the special case with $z_1 = z_2 = z$, we obtain
\begin{align}
&\frac{1}{N/2} { \sum_{\mib{k}} }' G^{\rm AA}_{ \mib{k}} (z) G^{\rm AA}_{ \mib{k}} (z) 
 = \frac{ \zeta_{\rm B}(z)}{2 \zeta_{\rm A}(z) } [  F_1 ({\tilde z}) + F_2({\tilde z})  ]
 \label{eq_formula_1b} \\
&\frac{1}{N/2} { \sum_{\mib{k}} }' G^{\rm AB}_{ \mib{k}} (z) G^{\rm BA}_{ \mib{k}} (z)
 = - \frac{1}{2}  [  F_1 ({\tilde z}) - F_2({\tilde z})]
 \label{eq_formula_2b}
\end{align}
These relations can also be obtained via $z$-derivative of eqs. (\ref{eq_formula_1}) and (\ref{eq_formula_2}).

It is easy to confirm that these expressions reproduce the susceptibility in the original Brillouin zone.
In the normal state, the relation $\zeta _{\rm A} = \zeta_{\rm B} = \zeta$ is satisfied.
Using eqs. (\ref{eq_formula_1b}) and (\ref{eq_formula_2b}), the uniform and staggered susceptibility defined in eq. (\ref{eq_sublattice_sum}) is given by
\begin{align}
\chi ^{\rm unif} (z) = - F_2 \left( \zeta (z) \right) , \\
\chi ^{\rm stag} (z) = - F_1 \left( \zeta (z) \right) .
\end{align}
Thus, the susceptibilities without vertex functions can be calculated from the functions $F_1$ and $F_2$.

\label{lastpage}
\clearpage

\end{document}